\documentclass[letterpaper, 10 pt, conference]{ieeeconf}
\IEEEoverridecommandlockouts

\usepackage[colorlinks]{hyperref}
\hypersetup{
    colorlinks,
    linkcolor=black,
    citecolor=black,
    filecolor=magenta,
    urlcolor=cyan,
}

\usepackage{resizegather}

\usepackage{cite}
\usepackage{amsmath,amssymb,amsfonts,mathrsfs}
\usepackage{algorithmic}
\usepackage{graphicx}
\usepackage{textcomp}
\usepackage{xcolor}
\usepackage{tcolorbox}

\usepackage{tablefootnote}
\usepackage{threeparttable}
\usepackage{caption, subcaption}

\usepackage{tikz}
\usepackage{graphicx}
\usepackage{tkz-euclide}

\usetikzlibrary{positioning}
\usetikzlibrary{shapes}
\usetikzlibrary{shapes.misc}
\usetikzlibrary{shapes.geometric}
\usetikzlibrary{plotmarks}
\usetikzlibrary{intersections}
\usetikzlibrary{calc}
\usetikzlibrary{fit}
\usetikzlibrary{patterns,tikzmark}
\usetikzlibrary{matrix,decorations.pathreplacing,calc}

\tikzset{cross/.style={cross out, draw, 
         minimum size=2*(#1-\pgflinewidth), 
         inner sep=0pt, outer sep=0pt}}

\DeclareMathOperator*{\argmin}{arg\,min}

\newcommand{\state}[0]{x}
\newcommand{\prel}[0]{p_{\rm{rel}}}
\newcommand{\vrel}[0]{v_{\rm{rel}}}
\newcommand{\vt}[0]{V_{\rm{T}}}
\newcommand{\vtdot}[0]{\dot{V}_{\rm{T}}}

\newcommand{\vreldot}[0]{\dot{v}_{\rm{rel}}}

\newcommand{\at}[0]{A_{\rm{T}}}

\newcommand{\norm}[1]{\left\lVert#1\right\rVert}

\newcommand{\vertiii}[1]{{\left\vert\kern-0.25ex\left\vert\kern-0.25ex\left\vert #1 
    \right\vert\kern-0.25ex\right\vert\kern-0.25ex\right\vert}}

\newtheorem{theorem}{Theorem}

\newtheorem{definition}{Definition}

\makeatletter
\renewcommand{\fps@figure}{htp}
\renewcommand{\fps@table}{htp}
\makeatother

\usepackage{eso-pic}
\newcommand\AtPageUpperMyright[1]{\AtPageUpperLeft{
 \put(\LenToUnit{0.5\paperwidth},\LenToUnit{-1cm}){
     \parbox{0.5\textwidth}{\raggedleft\fontsize{9}{11}\selectfont #1}}
 }}
\newcommand{\conf}[1]{
\AddToShipoutPictureBG*{
\AtPageUpperMyright{#1}
}
}

\def\BibTeX{{\rm B\kern-.05em{\sc i\kern-.025em b}\kern-.08em
    T\kern-.1667em\lower.7ex\hbox{E}\kern-.125emX}}
    
\begin{document}

\title{Real Time Safety of Fixed-wing UAVs using Collision Cone Control Barrier Functions}
\conf{\textbf{$8^{th}$ Cyber Physical System Symposium (CyPhySS), 2024. \\
25-27 July, 2024.}} 

\author{Aryan Agarwal*$^{1}$,  Ravi Agrawal*$^{2}$, Manan Tayal*$^{3}$, Pushpak Jagtap$^{3}$, Shishir Kolathaya$^{3}$.
\thanks{$^{1}$Indian Institute of Technology Kharagpur.
{\tt\scriptsize}
}%
\thanks{$^{2}$VNIT, Nagpur.
{\tt\scriptsize}
}%
\thanks{$^{3}$Cyber-Physical Systems, Indian Institute of Science (IISc), Bengaluru.
{\tt\scriptsize \{manantayal, shishirk\}@iisc.ac.in}
.
}%
\thanks{* denotes equal contribution. This work was done when Ravi and Aryan were interns at IISc, Bengaluru.
}
}

\maketitle
\begin{abstract}
Fixed-wing UAVs have transformed the transportation system with their high flight speed and long endurance, yet their safe operation in increasingly cluttered environments depends heavily on effective collision avoidance techniques. This paper presents a novel method for safely navigating an aircraft along a desired route while avoiding moving obstacles. We utilize a class of control barrier functions (CBFs) based on collision cones to ensure the relative velocity between the aircraft and the obstacle consistently avoids a cone of vectors that might lead to a collision. By demonstrating that the proposed constraint is a valid CBF for the aircraft, we can leverage its real-time implementation via Quadratic Programs (QPs), termed the CBF-QPs. Validation includes simulating control law along trajectories, showing effectiveness in both static and moving obstacle scenarios. 
\end{abstract}


\section{Introduction}
\label{section: Introduction}
Fixed-wing UAVs have revolutionized the transportation system with their high flight speeds and long endurance. However, their safe operation in increasingly cluttered environments poses a significant challenge, as effective collision avoidance techniques are crucial. Traditional control systems for safety-critical applications should ideally come with guarantees of safe operation under specified conditions. To address these challenges, run-time assurance (RTA) systems have been developed as supplementary modules for aircraft flight controllers, intervening to prevent safety violations.

Despite these advancements, designing effective control laws for RTA remains an unresolved issue. In response to this problem, we present a novel method for safely navigating an aircraft along a desired route while avoiding moving obstacles. Our approach utilizes a class of control barrier functions (CBFs) \cite{Ames_2017} based on collision cones to ensure that the relative velocity between the aircraft and obstacles consistently avoids vectors that might lead to a collision \cite{tayal2023control,goswami2023collision}. By demonstrating that the proposed constraint is a valid CBF for the aircraft, we enable its real-time implementation via Quadratic Programs (QPs), termed CBF-QPs. We validate our approach by simulating control laws along trajectories, showing its effectiveness in both static and moving obstacle scenarios.

Control systems for safety-critical applications should ideally come with guarantees of safe operation under specified conditions. Recently, run-time assurance (RTA) systems have been developed as supplementary modules for aircraft flight controllers to prevent safety violations. Designing effective control laws for RTA remains a challenging and unresolved issue.

\begin{figure}[t]
    \centering
    \includegraphics[width=1.0\linewidth]{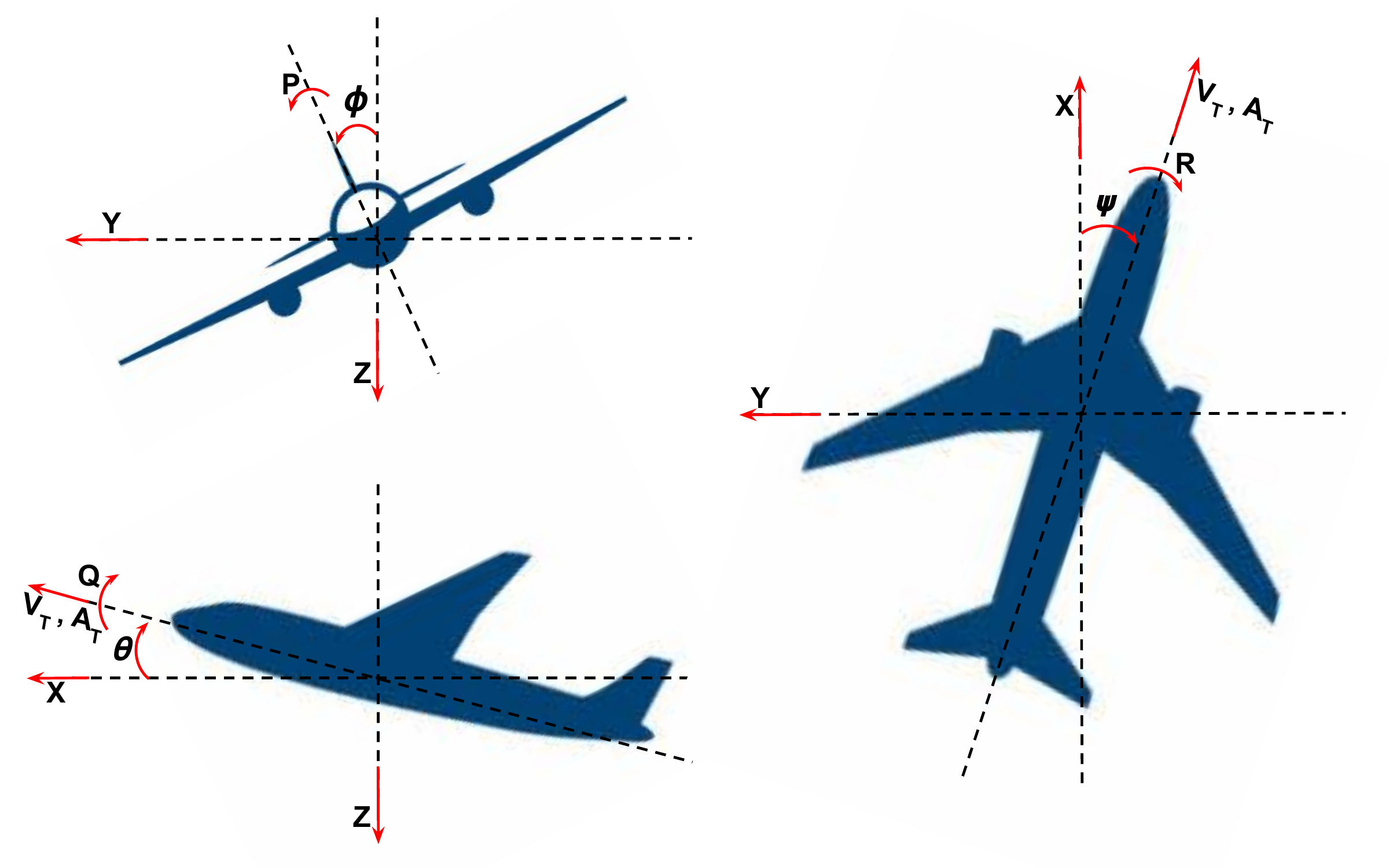}
\caption{Aircraft's linear and angular velocities defined in body-fixed frame. (X,Y,Z) represents earth-fixed frame.}
\label{fig:models}
\end{figure}

The rest of this paper is organized as follows. Preliminaries explaining the fixed-wing UAV model, the concept of control barrier functions (CBFs), collision cone CBFs, and controller design are introduced in section \ref{section: Background}. The application of the above CBFs on the fixed-wing aircraft to avoid obstacles modelled as spheres is discussed in section \ref{section: Safety Guarantee}. The simulation setup and comparative results will be discussed in section \ref{section: Simulation Results}. Finally, we present our concluding remarks in section \ref{section: Conclusions}.

\section{Preliminaries}
\label{section: Background}
In this section, first, we will describe the kinematics of fixed-wing UAV. Next, we will formally introduce Control Barrier Functions (CBFs) and their importance in providing formal safety guarantees in safety-critical applications. Finally, we will introduce  Collision Cone Control Barrier Function (C3BF) approach.

\subsection{Control barrier functions (CBFs)}
Here, we formally introduce Control Barrier Functions (CBFs) and their applications in the context of safety. 
%
Given the aircraft model \eqref{eqn:kinematic_model}, we have the nonlinear model in the control affine form:
\begin{equation}
	\dot{\state} = f(\state) + g(\state)u
	\label{eqn: affine control system}
\end{equation}
where $\state \in \mathcal{D} \subseteq \mathbb{R}^n$ is the state of system, and $u \in \mathbb{U} \subseteq \mathbb{R}^m$ the input for the system. Assume that the functions $f: \mathbb{R}^n \rightarrow \mathbb{R}^n$ and $g: \mathbb{R}^n \rightarrow \mathbb{R}^{n \times m}$ are continuously differentiable. Specific formulation of $f,g$ for the aircraft were described in \eqref{eqn:kinematic_model}. Given a Lipschitz continuous control law $u = k(\state)$, the resulting closed loop system $\dot{\state} = f_{cl}(\state) = f(\state) + g(\state)k(\state)$ yields a solution $\state(t)$, with initial condition $\state(0) = \state_0$.
%
Consider a set $\mathcal{C}$ defined as the \textit{super-level set} of a continuously differentiable function $h:\mathcal{D}\subseteq \mathbb{R}^n \rightarrow \mathbb{R}$ yielding,
\begin{align}
\label{eq:setc1}
	\mathcal{C} & = \{ \state \in \mathcal{D} \subset \mathbb{R}^n : h(\state) \geq 0\} \\
\label{eq:setc2}
	\partial\mathcal{C} & = \{ \state \in \mathcal{D} \subset \mathbb{R}^n : h(\state) = 0\}\\
\label{eq:setc3}
	\text{Int}\left(\mathcal{C}\right) & = \{ \state \in \mathcal{D} \subset \mathbb{R}^n : h(\state) > 0\}
\end{align}
It is assumed that $\text{Int}\left(\mathcal{C}\right)$ is non-empty and $\mathcal{C}$ has no isolated points, i.e. $\text{Int}\left(\mathcal{C}\right) \neq \phi$ and $\overline{\text{Int}\left(\mathcal{C}\right)} = \mathcal{C}$. 
The system is safe w.r.t. the control law $u = k(\state)$ if
	$\forall \: \state(0) \in \mathcal{C} \implies \state(t) \in \mathcal{C}$ $\forall t \geq 0$.
We can mathematically verify if the controller $k(\state)$ is safeguarding or not by using Control Barrier Functions (CBFs), which is defined next.

\begin{definition}[Control barrier function (CBF)]{\it
\label{definition: CBF definition}
Given the set $\mathcal{C}$ defined by \eqref{eq:setc1}-\eqref{eq:setc3}, with $\frac{\partial h}{\partial \state}(\state) \neq 0\; \forall \state \in \partial \mathcal{C}$, the function $h$ is called the control barrier function (CBF) defined on the set $\mathcal{D}$, if there exists an extended \textit{class} $\mathcal{K}$ function $\kappa$ such that for all $\state \in \mathcal{D}$:
\begin{equation}
\begin{aligned}
    \underbrace{\text{sup}}_{ u \in \mathbb{U}}\! \left[\underbrace{\mathcal{L}_{f} h(\state) + \mathcal{L}_g h(\state)u} \iffalse+ \frac{\partial h}{\partial t}\fi_{\dot{h}\left(\state, u\right)} \! + \kappa\left(h(\state)\right)\right] \! \geq \! 0
\end{aligned}
\end{equation}
where $\mathcal{L}_{f} h(\state) = \frac{\partial h}{\partial \state}f(\state)$ and $\mathcal{L}_{g} h(\state)= \frac{\partial h}{\partial \state}g(\state)$ are the Lie derivatives. 
}
\end{definition}

Given this definition of a CBF, we know from \cite{Ames_2017} and \cite{8796030} that any Lipschitz continuous control law $k(\state)$ satisfying the inequality: $\dot{h} + \kappa( h )\geq 0$ ensures safety of $\mathcal{C}$ if $x(0)\in \mathcal{C}$, and asymptotic convergence to $\mathcal{C}$ if $x(0)$ is outside of $\mathcal{C}$. 

\subsection{Safety Filter Design}
\label{subsection: safe_controller}
Having described the CBF, we can now describe the Quadratic Programming (QP) formulation of CBFs. CBFs act as \textit{safety filters} which take the desired input $u_{des}(\state,t)$ and modify this input in a minimal way: 
\begin{equation}
\begin{aligned}
\label{eqn:CBF_QP}
    u^{*}(x,t) &= \argmin_{u \in \mathbb{U} \subseteq \mathbb{R}^m} \norm{u - u_{des}(x,t)}^2\\
    \quad & \textrm{s.t. } \mathcal{L}_f h(x) + \mathcal{L}_g h(x)u + \kappa \left(h(x)\right) \geq 0\\
\end{aligned}
\end{equation}
This is called the Control Barrier Function based Quadratic Program (CBF-QP). The explicit form of the CBF-QP control $u^{*}$ can be obtained by solving the above optimization problem using KKT conditions:
\begin{equation}
\begin{aligned}
\label{eqn: CBF QP KKT1}
u^{*}(x,t) &= u_{des}(x,t) + u_{safe}(x,t)\\
\end{aligned}
\end{equation}
where $u_{safe}(x,t)$ is given by
\begin{multline}\label{eq:CBF-QP}
    u_{safe}(x, t) \!=\!
    \begin{cases}
        0 & \text{for } \psi(x, t) \geq 0 \\
        -\frac{\mathcal{L}_{g}h(x)^T \psi(x, t)}{\mathcal{L}_{g}h(x)\mathcal{L}_{g}h(x)^T} & \text{for } \psi(x, t) < 0
    \end{cases}
\end{multline}
where $\psi (x,t) := \dot{h}\left(x, u_{ref}(x, t)\right) + \kappa \left(h(x)\right)$. The sign change of $\psi$ yields a switching type of a control law.

\subsection{Collision Cone CBF (C3BF) candidate for fixed wing UAVs}
\label{subsection: C3BF}
\begin{figure}[t]
    \includegraphics[width=1\linewidth]{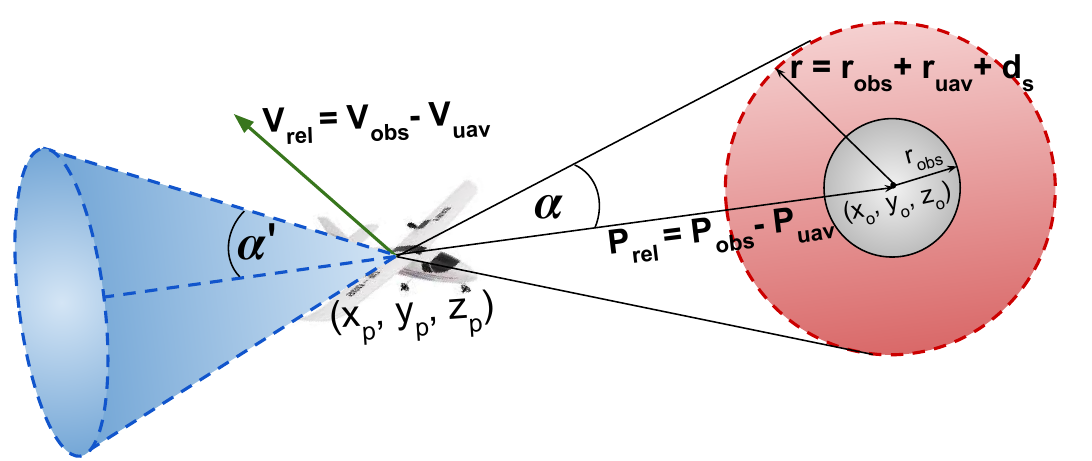}
\caption{\textbf{3D CBF} candidate: The dimensions of the obstacle are comparable to each other, it can be assumed as a sphere}
\label{fig:3D CBF}
\end{figure}

We now formally introduce the proposed CBF candidate for fixed wing UAVs. Let us assume that the obstacle is centered at $(x_o(t), y_o(t), z_o(t))$ having maximum dimension of $r_{obs}$. We assume that $x_o(t), y_o(t), z_o(t)$ are differentiable and their derivatives are piece-wise constants. $r_{uav}$ represents the maximum radius of the sphere that circumscribes the UAV, $d_s$ is the minimum distance to be maintained between the UAV and the obstacle.
The proposed approach combines the idea of potential unsafe directions given by collision cone (Fig. \ref{fig:3D CBF}) as an unsafe set to formulate a CBF as in \cite{C3BF}.
Consider the following CBF candidate:
\begin{equation}
    h(\state, t) = < \prel, \vrel> + \| \prel\|\| \vrel\|\cos\alpha ,
    \label{eqn:CC-CBF}
\end{equation}
where $\prel$ is the relative position vector between the body center of the aircraft and the center of the obstacle, $\vrel$ is the relative velocity, $<\cdot , \cdot>$ is the dot product of 2 vectors and $\alpha$ is the half angle of the cone, the expression of $\cos\alpha$ is given by $\frac{\sqrt{\|\prel\|^2 - r^2}}{\|\prel\|}$ (see Fig. \ref{fig:3D CBF}). Precise mathematical definitions for $\prel, \vrel$ will be given in the next section. The proposed constraint simply ensures that the angle between $\prel, \vrel$ is less than $180^\circ - \alpha$.

In \cite{C3BF,tayal2023control,goswami2023collision,C3BF_tac}, it was shown that the proposed candidate \eqref{eqn:CC-CBF} is valid CBF for wheeled mobile robots, i.e., the unicycle and bicycle, and for quadrotors. With this result, CBF-QPs were constructed that yielded collision-avoiding behaviors in these models. We aim to extend this to the class of fixed wing UAVs. 

\section{Collision Cone CBFs on Fixed Wing UAV}
\label{section: Safety Guarantee}
Having described the Collision Cone CBF candidate, we will see their application on fixed wing UAVs in this section. We consider our CBF candidate in its naive form and one extended with a backstepping-based approach.

\subsection{Fixed Wing UAV model}
In this work, we adapt the \textit{3D Dubins kinematic model} \cite{molnar2024collision}, that describes the motion of fixed-wing aircraft. Kinematic models are more generalised and are invariant with changes in inertia or vechile dimensions, thus readily used for navigation and path planning tasks. The state of an aircraft in Dubins model is described by $x = [x_p, y_p, z_p, \phi, \theta, \psi, V_T]$.

\begin{equation}
\label{eqn:kinematic_model}
	\underbrace{\begin{bmatrix}
		\dot{x}_p \\
		\dot{y}_p \\
            \dot{z}_p \\
            \\
		\dot{\phi} \\
            \dot{\theta} \\
            \dot{\psi} \\
            \\
            \vtdot
	\end{bmatrix}}_{\dot{\state}}
	=
	\underbrace{\begin{bmatrix}
		\vt c\theta c\psi \\
		\vt c\theta s\psi \\
            -\vt s\theta \\
            \\
            \frac{g}{\vt}\hspace{-4pt}
            \begin{bmatrix}
                s\phi c\phi s\theta \\
                -(s\phi)^{2} c\theta \\
                s\phi c\phi
            \end{bmatrix}\\
            \\
            0
	\end{bmatrix}}_{f(\state)}
	+
	\underbrace{\begin{bmatrix}
		\\
            \begin{bmatrix}
                0 & 0 & 0 \\
                0 & 0 & 0 \\
                0 & 0 & 0 
            \end{bmatrix}\\
            \\
            \begin{bmatrix}
                0 & 1 & s\phi t\theta \\
                0 & 0 & c\phi \\
                0 & 0 & \frac{s\phi}{c\theta} 
            \end{bmatrix}\\
            \\
            \begin{bmatrix}
                1 & 0 & 0 
            \end{bmatrix}\\
	\end{bmatrix}}_{g(\state)}
	\underbrace{\begin{bmatrix}
		A_T \\
		P \\
            Q
	\end{bmatrix}}_{u}
\end{equation}

$x_p$, $y_p$, and $z_p$ denote the coordinates of the aircraft’s center of gravity in an earth-fixed frame. $\phi$, $\theta$ and $\psi$ represents the (roll, pitch \& yaw) orientation of the aircraft.  (see Fig. \ref{fig:models}). P, Q, R denote the respective angular velocities of the aircraft in the body frame, $V_T$ is the total longitudinal speed of the aircraft and $A_T$ is the longitudinal acceleration of the aircraft. $g$ is the gravitational acceleration constant.

\subsection{Naive C3BF candidate}
\label{section: 3D-CBF}
We first obtain the relative position and velocity vectors between the body centre of the fixed wing aircraft and the obstacle:
\begin{align}\label{eq:pos-vec-3D}
    \prel := \begin{bmatrix}
        x_o \\
        y_o \\
        z_o
    \end{bmatrix}
    -
    \begin{bmatrix}
        x_p \\
        y_p \\
        z_p
    \end{bmatrix}
\end{align}
$x_o, y_o, z_o$ represents the obstacle location as a function of time. Also, since the obstacles are of constant velocity, we have $\Ddot{x}_o= \Ddot{y}_o= \Ddot{z}_o = 0$. We obtain its relative velocity as
\begin{align}\label{eq:vel-vec-3D}
    \vrel := \dot{p}_{rel}
\end{align}

Now, we calculate the $\vreldot$ term which contains our inputs, i.e, $(\at, P, Q)$, as follows:
\begin{equation}
    \vreldot := -
    \begin{bmatrix}
        c\psi c\theta & 0 & -\vt (s\phi s\psi + s\theta c\phi c\psi) \\
        s\psi c\theta & 0 & \vt (s\phi c\psi - s\psi s\theta c\phi) \\
        -s\theta & 0 & -\vt c\phi c\theta
    \end{bmatrix}
    \begin{bmatrix}
        \at \\
        P \\
        Q
    \end{bmatrix} \nonumber \\
    + \textrm{additional terms.}
    \label{eqn:v_dot_with_u}
\end{equation}
Please note that the $\textrm{additional terms}$ in the above equation, refer to those terms that do not contain the input terms ($\at, P, Q$) and thus do not contribute to the calculation of $\mathcal{L}_g h$.

Having introduced the Collision Cone CBF candidate in \ref{subsection: C3BF}, we state the following result about its validity:

\begin{theorem}\label{thm:CC-CBF-3D}{\it
Given the fixed wing UAV model \eqref{eqn:kinematic_model}, the proposed CBF candidate \eqref{eqn:CC-CBF} with $\prel,\vrel$ defined by \eqref{eq:pos-vec-3D}, \eqref{eq:vel-vec-3D} is a valid CBF defined for the set $\mathcal{C}$.}
\end{theorem}

The way in which the controlled system works is shown in Fig. \ref{fig:cbf_comparision_results} and described in detail in \eqref{section: Simulation Results}. The main shortcoming of this CBF is that it is valid only for the set $\mathcal{C}$. Intuitively, this is due to the fact that $u_{safe}$ can only alter $\at$ and $Q$, but not $P$. This can be verified by seeing how $\mathcal{L}_g h$ is constructed using $\vreldot$ in \eqref{eqn:v_dot_with_u}:
\begin{align}
    \mathcal{L}_g h = \begin{bmatrix}
        < \xi(x,t), 
            \begin{bmatrix}
                c\psi c\theta \\
                s\psi c\theta \\
                -s\theta
            \end{bmatrix}>\\
        < \xi(x,t), 
            \begin{bmatrix}
                0 \\
                0 \\
                0
            \end{bmatrix}>\\
        < \xi(x,t), 
            \begin{bmatrix}
                -\vt (s\phi s\psi + s\theta c\phi c\psi) \\
                \vt (s\phi c\psi - s\psi s\theta c\phi) \\
                -\vt c\phi c\theta
            \end{bmatrix}>\\
    \end{bmatrix}^T,
\end{align}
where $\xi(x,t) = \prel + \vrel \sqrt{\|\prel\|^2 - r^2}/\|\vrel\|$

This shows that the coefficient of $P$ in $\mathcal{L}_g h(x) u$ is 0, rendering input $P$ uncontrollable by the safety filter.

\subsection{C3BF Candidate with Backstepping}
\label{section: back-CBF}

The safety set can be extended from $\mathcal{C}$ to $\mathcal{D}$ by considering a version of \eqref{eqn:CC-CBF} extended by backstepping. Using the defined kinematics \eqref{eqn:kinematic_model}, we can determine the desired roll rate $R_{des}$ from the desired acceleration $a_{des}$ based on the given trajectory. Since $R$ is a determined quantity, we \textit{backstep} from $R$ to find $P$, based on the method described in \cite{taylor2022safe}. The new CBF candidate is as follows:
\begin{equation}
    h(\state, t) = < \prel, \vrel> + \| \prel\|\| \vrel\|\cos\alpha - \frac{1}{2\lambda}(R_{des} - R)^2,
    \label{eqn:CC-CBF-back}
\end{equation}
where $\lambda > 0$ is a scaling constant.

We have the following result for the above CBF candidate extended through backstepping:
\begin{theorem}\label{thm:CC-CBF-back}{\it
Given the fixed wing UAV model \eqref{eqn:kinematic_model}, the proposed CBF candidate \eqref{eqn:CC-CBF-back} with $\prel,\vrel$ defined by \eqref{eq:pos-vec-3D}, \eqref{eq:vel-vec-3D} is a valid CBF defined for the set $\mathcal{D}$.}
\end{theorem}

The proof of the above result is beyond the scope of this paper and will be taken up in a future work. A descriptive comparison of \eqref{eqn:CC-CBF} and \eqref{eqn:CC-CBF-back} is given in the following section.

\section{Results and Discussions}
\label{section: Simulation Results}
In this section we describe the setup and parameters used for the Python simulation. Next, we do a comparative analysis of naive and backstepping based C3BFs, understanding how they affect the controlled system. Finally, we compare teh results from both our proposed C3BF candidates to an existing CBF  

\subsection{Simulation Setup}
\par We have validated the C3BF-QP based controller on fixed wing UAVs for Naive and Backstepped CBF cases. We have used an exponentially stable velocity tracking controller (Appendix B \cite{molnar2024collision}). Note that the choice of reference controller does not affect the validity of the CBFs proposed above, any valid velocity tracking controller can be used. 

For the class $\mathcal{K}$ function in the CBF inequality, we chose $\kappa(h) = \gamma h$, where $\gamma=1$. The scaling parameter in \eqref{eqn:CC-CBF-back} was chosen to be $\lambda = 10^{-4}$. The gravitational constant used in \eqref{eqn:kinematic_model} is $g = 9.81 m/s^2$. The collision radius was chosen to be $r = r_{obs} + r_{uav} + d_s = 100 m$, where $r_{obs}$ is the collision radius of the obstacle, $r_{uav}$ is the collision radius of the controlled aircraft and $d_s$ is a safety measure.

\subsection{C3BF - Naive vs Backstepped}
The way in which the controlled system works is layered with a base trajectory tracking controller with a C3BF-QP safety layer for collision avoidance. The base controller $u_{des}$ is designed to make the aircraft follow a given trajectory. When an obstacle, static or having constant velocity, intersects the aircraft's trajectory with impending collision, the C3BF-QP safety filter starts to alter the $u_{des}$ in order to obtain a net control policy which ensures that the aircraft remains `safe', i.e. avoids the collision. 

In the case of the naive C3BF \eqref{eqn:CC-CBF}, $u_{safe}$ derived from \eqref{eqn:CBF_QP} alters the \textit{desired} $\at$ and $Q$, provided by $u_{des}$ such that the new control policy $u_{des} + u_{safe}$ avoids the obstacle on the path. The aircraft only pitches up or down in order to avoid the obstacle while slowing down or speeding up accordingly.

\begin{figure}
    \centering
    \begin{subfigure}[b]{0.3\textwidth}
        \includegraphics[width=\textwidth]{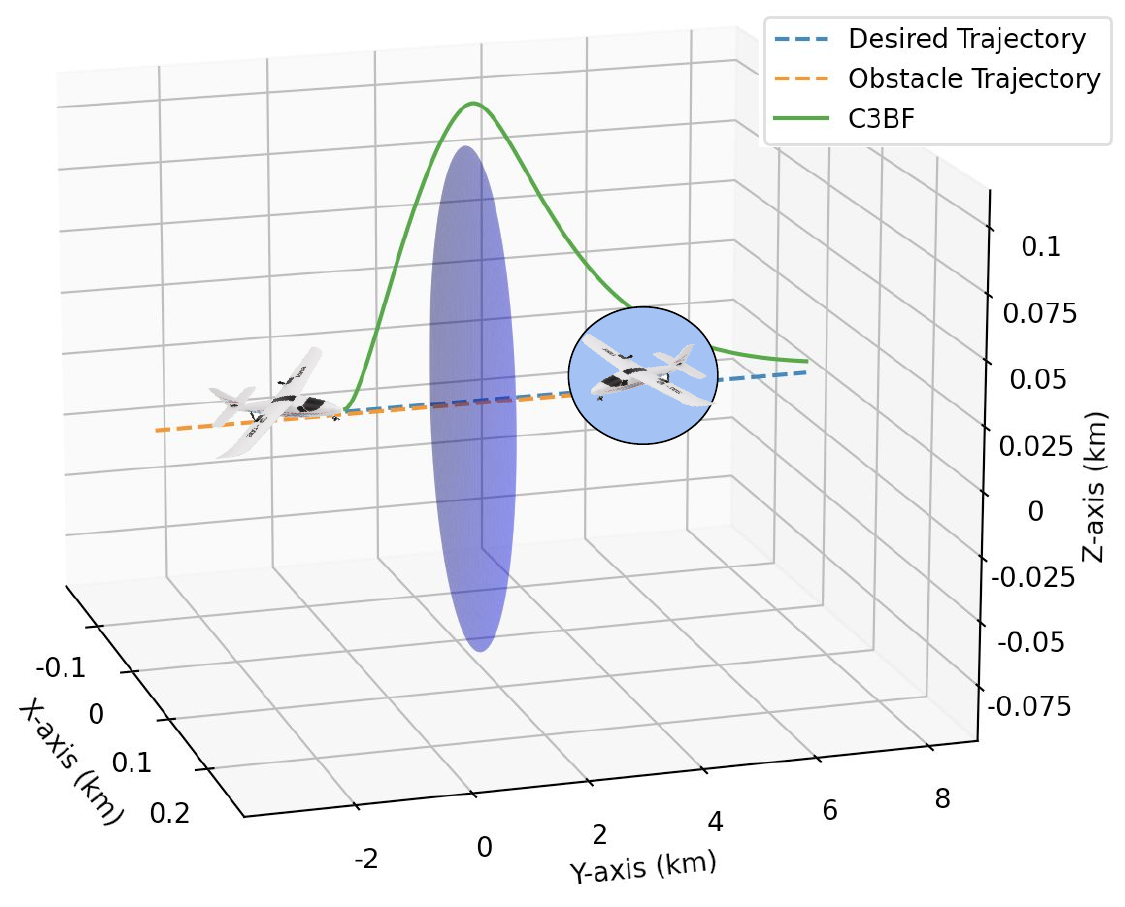}
        \caption{}
    \end{subfigure}
    
    \begin{subfigure}[b]{0.3\textwidth}
        \includegraphics[width=\textwidth]{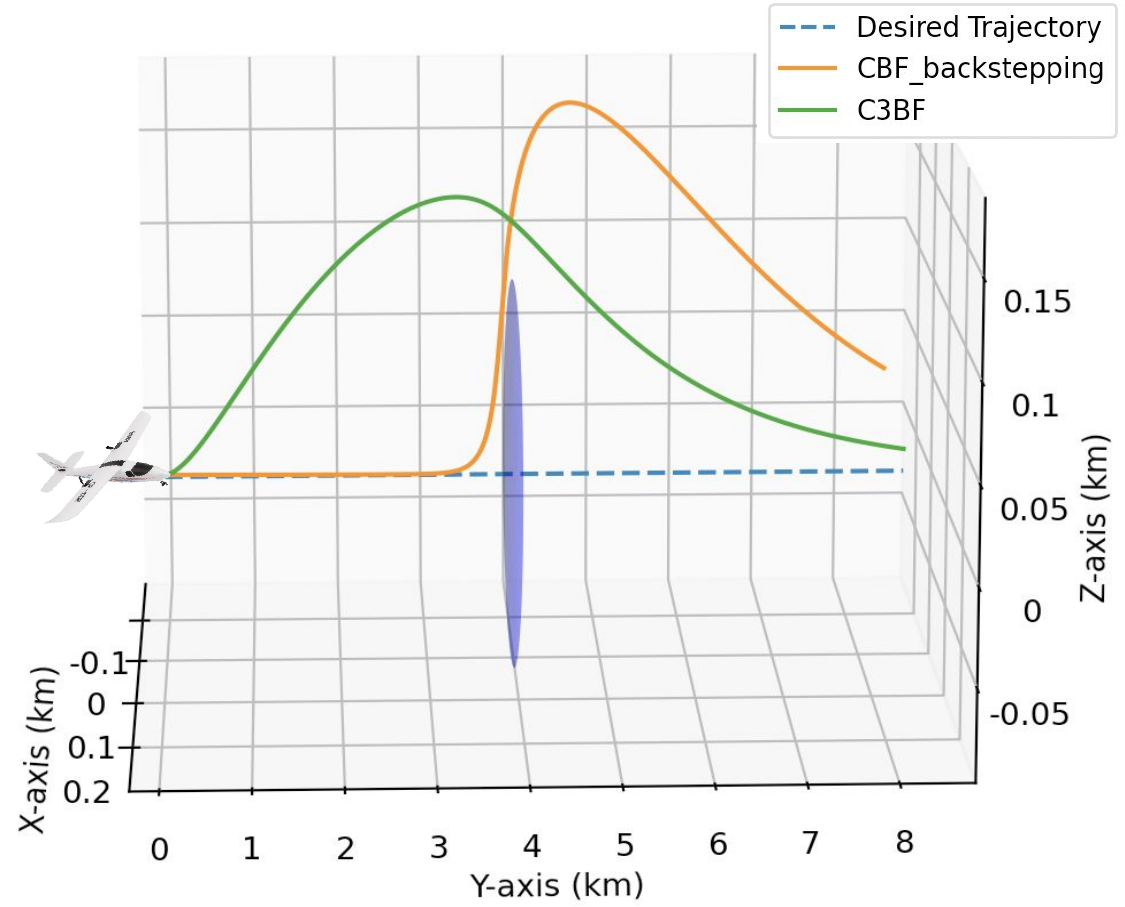}
        \caption{}
    \end{subfigure}
    
    \begin{subfigure}[b]{0.3\textwidth}
        \includegraphics[width=\textwidth]{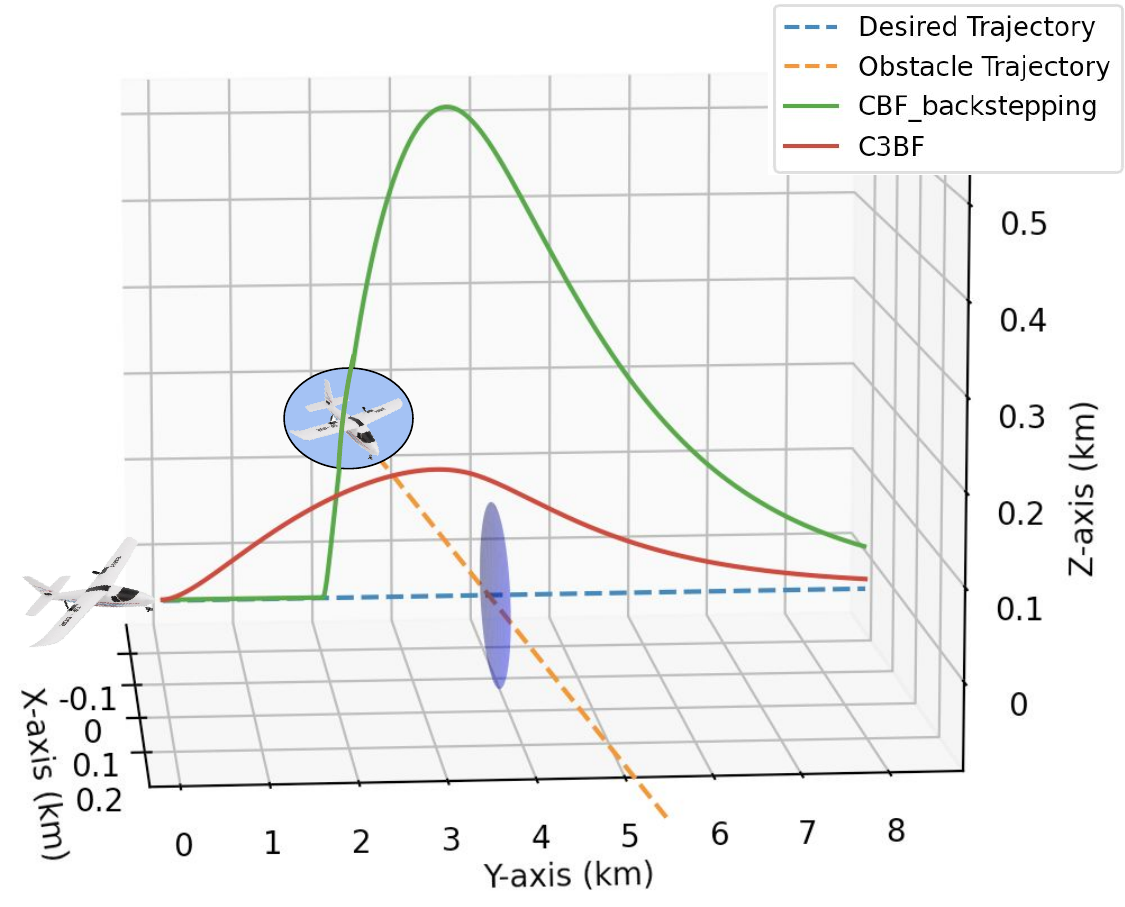}
        \caption{}
    \end{subfigure}
    \caption{Aircraft's behaviour following C3BF: Obstacle incoming head-on to the aircraft (a), Comparision of aircraft's behaviour following C3BF and CBF in \cite{molnar2024collision} for a static obstacle (b) and for constant velocity obstacle (c). In all three cases, the collision radius $r = 100m$.}
    \label{fig:cbf_comparision_results}
\end{figure}

In case of obstacles which cannot be flown over, e.g. pillars, geofences, etc., $u_{safe}$ of \eqref{eqn:CC-CBF} is unable to maneuver the aircraft sideways, and can only make the aircraft stop in order to ensure safety. Hence the C3BF described by \eqref{eqn:CC-CBF-back} can generate $u_{safe}$ from \eqref{eq:CBF-QP} which brings change in all of $\at$, $P$ and $Q$, making the new CBF valid for the entire set $\mathcal{D}$. This also ensures stability of set $\mathcal{C}$ with respect to set $\mathcal{D}$. However, for the case of an obstacle with a finite 3D collision radius, following a constant velocity trajectory or static in the path of the aircraft, \eqref{eqn:CC-CBF} and \eqref{eqn:CC-CBF-back} produce $u_{safe}$ which behave identically, since the obstacle can always be flown over.

\subsection{C3BF vs existing CBF}
Comparing the action of the CBF proposed in \cite{molnar2024collision} to the C3BFs proposed in \eqref{eqn:CC-CBF} and \eqref{eqn:CC-CBF-back}, (Fig. \ref{fig:cbf_comparision_results} b, c) we see that the C3BFs safety filter start acting on the system much earlier allowing for a smoother path. Also the action of the CBF-QP in \cite{molnar2024collision} is more conservative, taking a path with a large deviation from the trajectory, whereas the action of C3BF-QPs from \eqref{eqn:CC-CBF} and \eqref{eqn:CC-CBF-back} chooses a path with minimal deviation while ensuring safety. This also shows that C3BFs create control policies which require a lower effort.




\section{Conclusions}
\label{section: Conclusions}
We have shown that Collision Cone CBF works for fixed wing UAVs. However, the point to be noted is that this work is under progress and we are yet to compare the results of Naive C3BF with C3BF with Backstepping for cases which set them apart in simulations. Moreover, we also plan to extend this analysis to collision avoidance with long obstacles like trees and towers, while providing theoretical guarantees. 

\label{section: References}
\bibliographystyle{IEEEtran}
\bibliography{references.bib}

\begin{thebibliography}{1}
\providecommand{\url}[1]{#1}
\csname url@samestyle\endcsname
\providecommand{\newblock}{\relax}
\providecommand{\bibinfo}[2]{#2}
\providecommand{\BIBentrySTDinterwordspacing}{\spaceskip=0pt\relax}
\providecommand{\BIBentryALTinterwordstretchfactor}{4}
\providecommand{\BIBentryALTinterwordspacing}{\spaceskip=\fontdimen2\font plus
\BIBentryALTinterwordstretchfactor\fontdimen3\font minus \fontdimen4\font\relax}
\providecommand{\BIBforeignlanguage}[2]{{%
\expandafter\ifx\csname l@#1\endcsname\relax
\typeout{** WARNING: IEEEtran.bst: No hyphenation pattern has been}%
\typeout{** loaded for the language `#1'. Using the pattern for}%
\typeout{** the default language instead.}%
\else
\language=\csname l@#1\endcsname
\fi
#2}}
\providecommand{\BIBdecl}{\relax}
\BIBdecl

\bibitem{Ames_2017}
\BIBentryALTinterwordspacing
A.~D. Ames, X.~Xu, J.~W. Grizzle, and P.~Tabuada, ``Control barrier function based quadratic programs for safety critical systems,'' \emph{{IEEE} Transactions on Automatic Control}, vol.~62, no.~8, pp. 3861--3876, aug 2017. [Online]. Available: \url{https://doi.org/10.1109%2Ftac.2016.2638961}
\BIBentrySTDinterwordspacing

\bibitem{tayal2023control}
M.~Tayal, R.~Singh, J.~Keshavan, and S.~Kolathaya, ``Control barrier functions in dynamic uavs for kinematic obstacle avoidance: a collision cone approach,'' \emph{arXiv preprint arXiv:2303.15871}, 2023.

\bibitem{goswami2023collision}
B.~G. Goswami, M.~Tayal, K.~Rajgopal, P.~Jagtap, and S.~Kolathaya, ``Collision cone control barrier functions: Experimental validation on ugvs for kinematic obstacle avoidance,'' \emph{arXiv preprint arXiv:2310.10839}, 2023.

\bibitem{8796030}
A.~D. Ames, S.~Coogan, M.~Egerstedt, G.~Notomista, K.~Sreenath, and P.~Tabuada, ``Control barrier functions: Theory and applications,'' in \emph{2019 18th European Control Conference (ECC)}, 2019.

\bibitem{C3BF}
P.~Thontepu, B.~G. Goswami, M.~Tayal, N.~Singh, S.~S. P~I, S.~S. M~G, S.~Sundaram, V.~Katewa, and S.~Kolathaya, ``Collision cone control barrier functions for kinematic obstacle avoidance in ugvs,'' in \emph{2023 Ninth Indian Control Conference (ICC)}, 2023, pp. 293--298.

\bibitem{C3BF_tac}
M.~{Tayal}, B.~{Giri Goswami}, K.~{Rajgopal}, R.~{Singh}, T.~{Rao}, J.~{Keshavan}, P.~{Jagtap}, and S.~{Kolathaya}, ``{A Collision Cone Approach for Control Barrier Functions},'' \emph{arXiv e-prints}, p. arXiv:2403.07043, 2024.

\bibitem{molnar2024collision}
T.~G. Molnar, S.~K. Kannan, J.~Cunningham, K.~Dunlap, K.~L. Hobbs, and A.~D. Ames, ``Collision avoidance and geofencing for fixed-wing aircraft with control barrier functions,'' 2024.

\bibitem{taylor2022safe}
A.~J. Taylor, P.~Ong, T.~G. Molnar, and A.~D. Ames, ``Safe backstepping with control barrier functions,'' 2022.

\end{thebibliography}

\end{document}